\documentclass[review]{elsarticle}

\usepackage{lineno,hyperref,aas_macros,threeparttable,amssymb}
\modulolinenumbers[5]

\biboptions{authoryear}

\begin{document}

\begin{frontmatter}

\title{Fates of hydrous materials during planetesimal collisions}

\author[elsiad]{Shigeru Wakita\corref{corresponding}}
\author[elsiad]{Hidenori Genda}
\cortext[corresponding]{Corresponding author}
\address[elsiad]{Earth-Life Science Institute, Tokyo Institute of Technology 2-12-1 Ookayama, Meguro-ku, Tokyo 152-8550, Japan}

\begin{abstract}
Hydrous minerals are found on the surfaces of asteroids, but their origin is not clear.
If their origin is endogenic, the hydrous minerals that were formed in the inner part of a planetesimal
 (or parent body) should come out on to the surface without dehydration.
 If their origin is exogenic, the source of hydrous minerals accreting onto asteroids is needed.
Collisions in the asteroid belt would be related to both origins
because collisions excavate the surface and eject the materials.
However, the fate of hydrous minerals in large planetesimals during the collisional process has not been well investigated.
Here, we explore planetesimal collisions by using the iSALE-2D code, 
and investigate the effect of an impact for the target planetesimal containing hydrous minerals.
Our numerical results for the fiducial case (5 km/s of the impact velocity) show that hydrous minerals are slightly heated during the collisions.
This moderate heating indicates that they can avoid the dehydration reaction and keep their original composition.
Some hydrous minerals have larger velocity than the escape velocity of the collision system.
This means that hydrous minerals can escape from the planetesimal and support the theory of exogenic origin for the hydrous minerals on asteroids.
Meanwhile, the velocity of other hydrous minerals is smaller than the escape velocity of the system.
This also indicates the possibility of an endogenic origin for the hydrous minerals on asteroids.
Our results suggest that hydrous minerals on asteroids can be provided by planetesimal collisions.
\end{abstract}


\end{frontmatter}


\section{Introduction} \label{sec:int}
Asteroids can be classified into several types \citep[e.g.,][]{DeMeo:2015aa}.
Some studies report that C-type asteroids contain hydrous materials \citep[e.g.,][]{Takir:2015aa,Rivkin:2015aa}. 
Additionally there are reports of evidence of hydrogen on the dwarf planet Ceres \citep{DeSanctis:2016aa} and asteroid Vesta \citep{Palmer:2017aa} via the Dawn mission. 
Recent work suggests that hydrous minerals (and water ice) on Ceres would be the result of Ceres's current activity \citep{Carrozzo:2018aa,Raponi:2018aa}.
Essentially there would be two origins for the hydrous minerals found on the surfaces of asteroids \citep[][]{Russell:2015aa}.
One is an endogenic origin, which means that the hydrous minerals were originally located inside the asteroids (planetesimals) and excavated via impacts.
Excavation of hydrated minerals to the surface on an asteroid implies that the asteroid avoided catastrophic disruption and re-accretion, i.e., such an asteroid is primordial.
Another is exogenic, which is when the impactor (other asteroids or planetesimals) brings the hydrated materials.
Both origins require impact events, hence planetesimal collisions.
It is not well understood how the collisions have an effect on the hydrous materials in the planetesimals.
Thus, it is important to examine collisions to understand the hydrous materials on the surfaces of asteroids.

Hydrous minerals, such as serpentine and saponite, are also found in carbonaceous chondrite \citep[e.g.,][]{Krot:2015aa}.
The reflectance spectrum of asteroids indicates that some of them are similar to those of meteorites \citep{DeMeo:2015aa,Reddy:2015aa}.
Meteorites, which originate from asteroids, can hold information about the period when they were embedded in their parent bodies \citep[e.g.,][]{Davis:2014aa}.  
Hydrous minerals are products of aqueous alteration, which is thought to have occurred in their parent bodies (planetesimals). 
This alteration could be triggered by the decay heat of short-lived radionuclides (e.g., $^{26}$Al),
which is followed by thermal evolution of planetesimals \citep[e.g.,][]{Grimm:1989aa,Gail:2014aa,Wakita:2014aa}.
Hence, hydrous minerals found in meteorites originally formed inside planetesimals.
However, hydrous minerals may experience a dehydration reaction
and become dehydrated like the ones found in chondrite \citep[e.g.,][]{Nakamura:2005aa}.
When the surrounding temperature exceeds a critical temperature ($\sim$ 600 $^\circ$C) \citep{Lange:1982aa,Nozaki:2006aa,Nakato:2008aa},
the dehydration reaction occurs and hydrous minerals become dehydrated.
Internal heating of planetesimals is one heat source that can trigger the dehydration reaction \citep[e.g.,][]{Grimm:1989aa,Wakita:2014aa}, 
and the other is temperature increase during planetesimal collisions.
Indeed, the mineralogical studies on some carbonaceous chondrites indicate impact heating could be a heat source causing dehydration \citep[e.g.,][]{Nakamura:2005aa,Nakato:2008aa,Abreu:2013aa}. 
Parent bodies of most carbonaceous chondrites do not reach such a high temperature because their peak metamorphic temperature is much lower than 600 $^\circ$C \citep{Scott:2014aa}.
Due to this, internal heating that heats up the whole body is excluded and the possible heat source for the dehydration minerals is impact heating.
Therefore, it is important to consider the possibility fo dehydration reactions when investigating planetesimal collisions.

In this paper, we examine the fate of hydrous materials during planetesimal collisions,
focusing in particular on the occurrence of the dehydration reaction and the amount of dehydrated materials.
In \S \ref{sec:met}, we demonstrate how to perform numerical simulations of planetesimal collisions using iSALE-2D.
Our numerical results are given in \S \ref{sec:res}.
We discuss some numerical issues and implications for asteroids and meteorites in \S \ref{sec:dis}.
Our concluding remarks are shown in \S \ref{sec:con}.

\section{Methods}\label{sec:met}

In this study we examine planetesimal collisions, focusing especially on dehydration of hydrous minerals that originated in a large target planetesimal. 
We perform numerical simulations of planetesimal collisions 
to evaluate the increase of absolute specific entropy and ejected mass 
using the shock physics code iSALE-2D \citep{Wunnemann:2006aa}, the version of which is iSALE-Dellen. 
Here absolute specific entropy is the absolute entropy per unit mass, and absolute entropy is defined as zero at 0 K.
The iSALE was developed to model planetary impacts and cratering based on the SALE hydrocode \citep{Amsden:1980aa}. 
The code has been improved from SALE by including various equations of state (EOS), a strength model, and a porosity compaction model \citep{Melosh:1992aa, Ivanov:1997aa, Collins:2004aa, Wunnemann:2006aa, Collins:2016aa}.  
Parameters for material properties used in this study are summarized in Table \ref{tab:isale}.
Although we do not know the actual strength parameters for planetesimals, we used the typical values for these parameters
 that were used in previous papers \citep[e.g.,][]{Johnson:2015aa, Kurosawa:2018aa}.
According to \citet{Kurosawa:2018aa}, the undamaged frictional coefficient is the most important parameter among the strength parameters regarding the material heating. 
Therefore, we expect that our results do not change so much as long as this coefficient is $\sim$ 0.1.
The effect of porosity are not included in this study for simplicity,
but elastic and plastic behavior with damage and friction models are included.
We do not include the effect of gravity in most of our calculations, except for one model with central gravity.

\begin{threeparttable}
\caption{iSALE input parameters}
\begin{tabular}{ll}
\hline
\hline
Description & Values \\
\hline
Equation of state &  ANEOS \tnote{a}\\
Bulk material of impactor/target & dunite \& serpentine\\
Solidus temperature & 1373 K \tnote{b,c} ~~~\&~ 1098 K \tnote{c,d} \\
Simon approximation constant A \tnote{e} & 1520 MPa\\
Simon approximation exponent C \tnote{e} & 4.05 \\
Poisson's ratio \tnote{f} & 0.25 \\
Thermal softening parameter \tnote{g} & 1.1 \\
Strength model \tnote{h} & Rock \\
Cohesion (damaged) \tnote{g} & 0.01 MPa \\
Cohesion (undamaged) \tnote{g} & 5.07 MPa \\
Frictional coefficient (damaged) \tnote{g} & 0.63 \\
Frictional coefficient (undamaged) \tnote{g} & 1.58 \\
Strength at infinite pressure \tnote{g} & 3.26 GPa \\
Damage model \tnote{h} & Ivanov \\
Minimum failure strain &  $10^{-4}$ \\
Damage model constant &  $10^{-11}$ \\
Threshold pressure for damage model &  300 MPa \\
\hline
\end{tabular}
\begin{tablenotes}
\item[a] dunite \citet{Benz:1989aa}, serpentine \citet{Brookshaw:1998aa}
\item[b] \citet{Keil:1997aa}
\item[c] \citet{Davison:2016aa}
\item[d] \citet{Till:2012aa}
\item[e] \citet{Davison:2010aa}
\item[f] \citet{Ivanov:1997aa}
\item[g] \citet{Johnson:2015aa}
\item[h] \citet{Collins:2004aa}
\end{tablenotes}
\label{tab:isale}
\end{threeparttable}

We consider the collisions between an impactor with three different radii ($R_{\rm imp}$) 
and a target with a fixed radius of 100 km ($R_{\rm tar}$) for three different impact velocities ($v_{\rm imp}$), as listed in Table \ref{tab:par}.
For the impactor, we consider anhydrous minerals, 
while we consider a two-layer model with a hydrous core covered with an anhydrous layer for the target.
The temperature evolution of planetesimals (typically $\ge 50$ km) is governed by the decay heat of short-lived radionuclides, 
such as $^{26}$Al (half-life of 0.72 Myr) \citep[e.g.,][]{Miyamoto:1982aa}.
When planetesimals form a few Myrs after CAI formation,
the temperature of the inner part of the planetesimals increases beyond 0 $^\circ$C (273K; \citet{Grimm:1993aa}).
This causes aqueous alteration if the planetesimals contain water and/or ice \citep[e.g.,][]{Grimm:1989aa,Wakita:2011aa}.
Then, hydrous material would form in the inner part of planetesimals.
On the other hand, the surface layer stays cool due to radiative cooling.
Here we consider a hydrous core with a radius of $R_{\rm hyd}$ ($< R_{\rm tar}$).
Serpentine is one of the major products of aqueous alteration and found in carbonaceous chondrite \citep[e.g.,][]{Krot:2015aa}.
Therefore, we assume that a hydrous core is composed of serpentine in our simulations.
In our fiducial case, $R_{\rm hyd}$ is 90 km which corresponds to a water/rock mass ratio of 0.3 before the aqueous alteration.
We take dunite as the component of the impactor and assume an anhydrous layer of the target planetesimal.
This is because it is similar to ordinary chondrites \citep{Svetsov:2015aa}, and can represent anhydrous materials.
All of these materials are taken from ANEOS in iSALE-2D code (dunite \citep{Benz:1989aa}, serpentine \citep{Brookshaw:1998aa}).

\begin{threeparttable}[b]
\caption{Parameter sets for simulation runs}
\begin{tabular}{rlll}
\hline
\hline
Model & $v_{\rm imp}$ [km/s] & $R_{\rm imp}$ [km] & $R_{\rm hyd}$ [km]\\
\hline
a)\tnote{*} & 2.5 & 20 & 90 \\
b) & 2.5 & 10 & 90 \\
c) & 2.5 & 20 & 90 \\
d) & 2.5 & 40 & 90 \\
e)\tnote{*} & 5 & 20 & 90 \\
f) & 5 & 10 & 90 \\
g) & 5 & 20 & 90 \\
h) & 5 & 40 & 90 \\
i) & 5 & 40 & 70 \\
j)\tnote{$\dagger$} & 5 & 20 & 90 \\
k)\tnote{*} & 10 & 20 & 90 \\
l) & 10 & 10 & 90 \\
m) & 10 & 20 & 90 \\
n) & 10 & 40 & 90 \\
\hline
\end{tabular}
\begin{tablenotes}
\item[*] runs without material strength
\item[$\dagger$] run with central gravity
\end{tablenotes}
\label{tab:par}
\end{threeparttable}

Hydrous materials would lose their water through the dehydration reaction.
The dehydration reaction starts to occur at around 600 $^\circ$C, based on previous experimental works \citep{Lange:1982aa,Nozaki:2006aa,Nakato:2008aa}.
We chose 600 $^\circ$C as a threshold temperature to distinguish hydrous materials and anhydrous ones.
Since we use the temperature at the beginning of the dehydration reaction as a threshold temperature, 
this could result in an overestimation of the amount of dehydrated minerals and an underestimation of the surviving amount of hydrous minerals.
However, since our focus is on whether hydrous minerals can survive dehydration or not,
our choice of this threshold temperature is feasible in our current work.
As shown in \citet{Lange:1982aa}, dehydration does not depend on pressure.
The temperature of 600 $^\circ$C in serpentine corresponds to the absolute specific entropy ($S$) of $S_{\rm dehyd} \equiv$ 3.2143 kJ/K/kg, 
which is calculated from ANEOS tables in iSALE-2D.
In analysis of collisional simulations, once the $S$ of hydrous materials exceeds $S_{\rm dehyd}$, we define it as a dehydrated part.
We do not switch the EOS from serpentine to dunite in the collisional simulations.
This is because we focus on the occurrence of the dehydration reaction and what occurs after the dehydration reaction is irrelevant to this study.

\begin{figure}
\includegraphics[clip,scale=0.8]{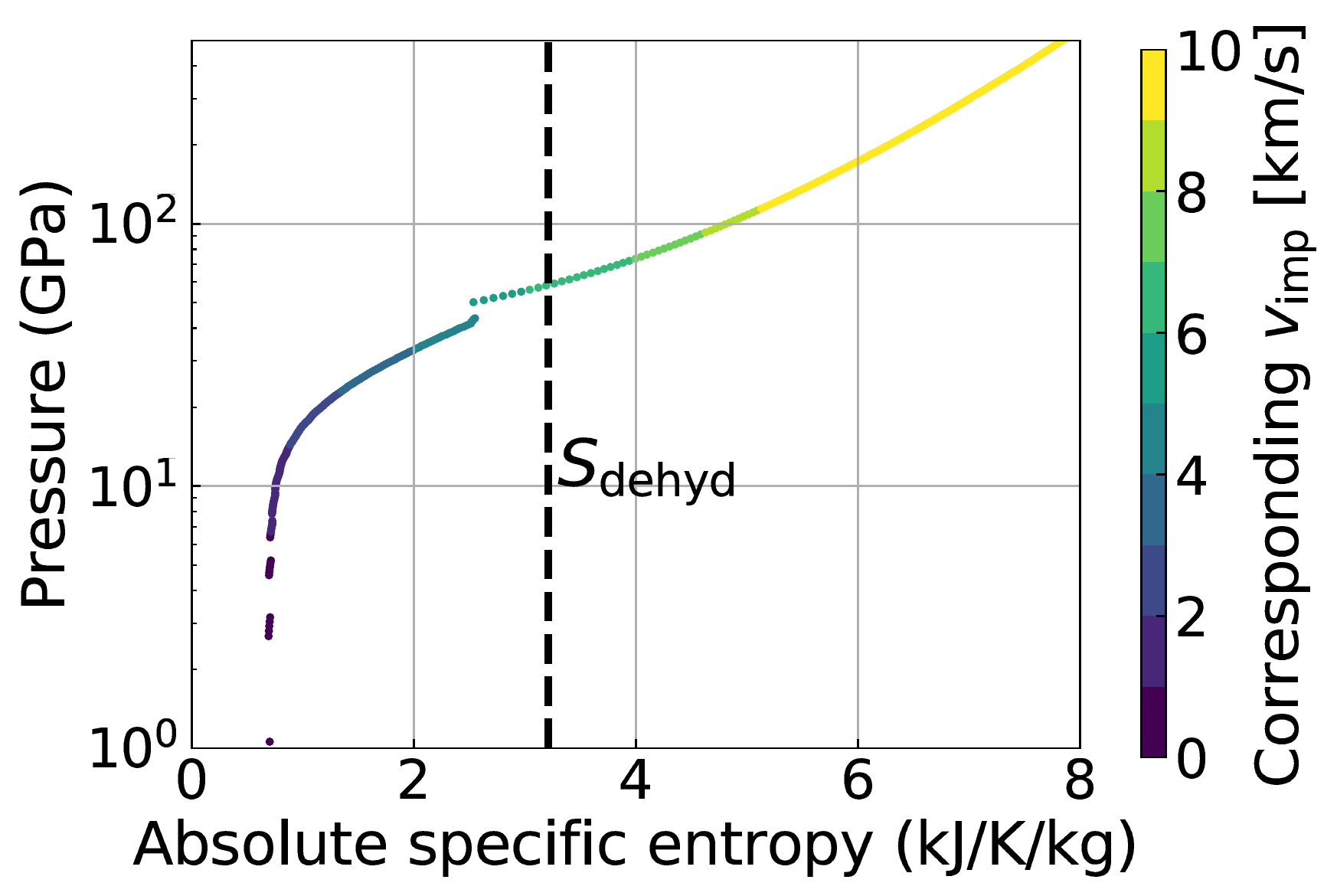}
\caption{Hugoniot curve for serpentine with the corresponding impact velocity $v_{\rm imp}$.
\label{fig:vimp}}
\end{figure}

Figure \ref{fig:vimp} shows the Hugoniot curve for serpentine derived from ANEOS table in iSALE code which is based on \citet{Brookshaw:1998aa}.
When the impact velocity exceeds 6 km/s, $S$ also exceeds $S_{\rm dehyd}$. 
Therefore, for a purely hydrodynamic case, the typical impact velocity ($\sim$ 5 km/s) of asteroids in the current main asteroid belt \citep{Farinella:1992aa,Bottke:1994aa}
cannot cause a significant dehydration reaction.
However, a recent study considering material strength \citep{Kurosawa:2018aa} reports that frictional heating during an impact increases enormously $S$.
Numerical simulations with the strength model are needed for precise evaluation of the dehydration reaction during planetesimal collisions.
Note that the dehydration reaction would be an endothermic reaction \citep{Wakita:2011aa}.
The main purpose of this study is to understand whether or not planetesimal collision would trigger the dehydration reaction.
Thus, we do not take into account the heat consumption in this work due to the dehydration reaction.

When the dehydration reaction occurs, it is not only dehydrated material (olivine) that is produced, but also water (H$_2$O).
When the vapor pressure of water $P_{\rm vap}$ exceeds lithostatic pressure $P_{\rm litho}$ and the tensile strength of rock $\tau$ (= 10 MPa; \citet{Cohen:2000aa}),
\begin{equation}
P_{\rm vap} > P_{\rm lith} + \tau ,
\end{equation}
venting of water occurs \citep{Grimm:1989aa,Wilson:1999aa,Cohen:2000aa}.
The vapor pressure of water is given by 
\begin{equation}
P_{\rm vap}(T) = P_0 \exp(-T_0/T),
\end{equation}
where $P_0$ = 4,700 MPa and $T_0$ = 4960 K \citep{Grimm:1989aa}.
At T = 600$^\circ$C, $P_{\rm vap}$ is about 150 MPa.

The lithostatic pressure $P_{\rm lith}$ at $r$ in the target planetesimal could be calculated as follows:
\begin{eqnarray}
P_{\rm lith}(r) = & \frac{4}{3} \pi G \left(\int^{R_{\rm tar}}_r \rho^2 r^\prime dr^\prime \right) \nonumber \\
= & \frac{4}{3} \pi G \left(\rho_{\rm dun}^2 R_{\rm tar}^2 - \rho_{\rm serp}^2 R_{\rm hyd}^2 +  \int^{R_{\rm hyd}}_r \rho_{\rm serp}^2 r^\prime dr^\prime \right),
\end{eqnarray}
where $G$ is the gravitational constant, $\rho_{\rm dun}$ = 3320 kg/m$^3$ is density of the anhydrous layer (dunite),
and $\rho_{\rm serp}$ = 2500 kg/m$^3$ is density of the hydrous core (serpentine).
At the center of planetesimals, $P_{\rm lith}(0)$ is about 20 MPa.
Thus, $P_{\rm lith}+\tau$ is 30 MPa at the center,
which means $P_{\rm vap}$ exceeds $P_{\rm lith} + \tau$ at any depth in the planetesimals with $R_{\rm tar}$ = 100 km once $T$ is above 600 $^\circ$C.
Therefore, we can expect that the water vapor produced via the dehydration reaction immediately escapes from the target planetesimals into space.

\section{Results} \label{sec:res}
In this section, the behavior of materials during planetesimal collisions is presented.
We focus in particular on entropy change during planetesimal collisions, which directly relates to dehydration.
We perform 13 simulations for planetesimal collisions between an impactor with $R_{\rm imp}$ and a target with $R_{\rm tar}$
containing a hydrous core with $R_{\rm hyd}$ for an impact velocity of $v_{\rm imp}$ (see Table \ref{tab:par}). 
Unless otherwise mentioned, we take the following parameter sets for the fiducial case:
$R_{\rm imp}$ = 20 km, $R_{\rm tar}$ = 100 km, $R_{\rm hyd}$ = 90 km, and $v_{\rm imp}$ = 5 km/s.
This impact velocity is a typical value in the current main asteroid belt \citep{Farinella:1992aa,Bottke:1994aa}.
We take -123$^\circ$C (150 K) as an initial temperature for impactor and target, and 40 cells per projectile radius for the typical numerical resolution.

\subsection{Fiducial case}

\begin{figure}
\includegraphics[clip,scale=0.8]{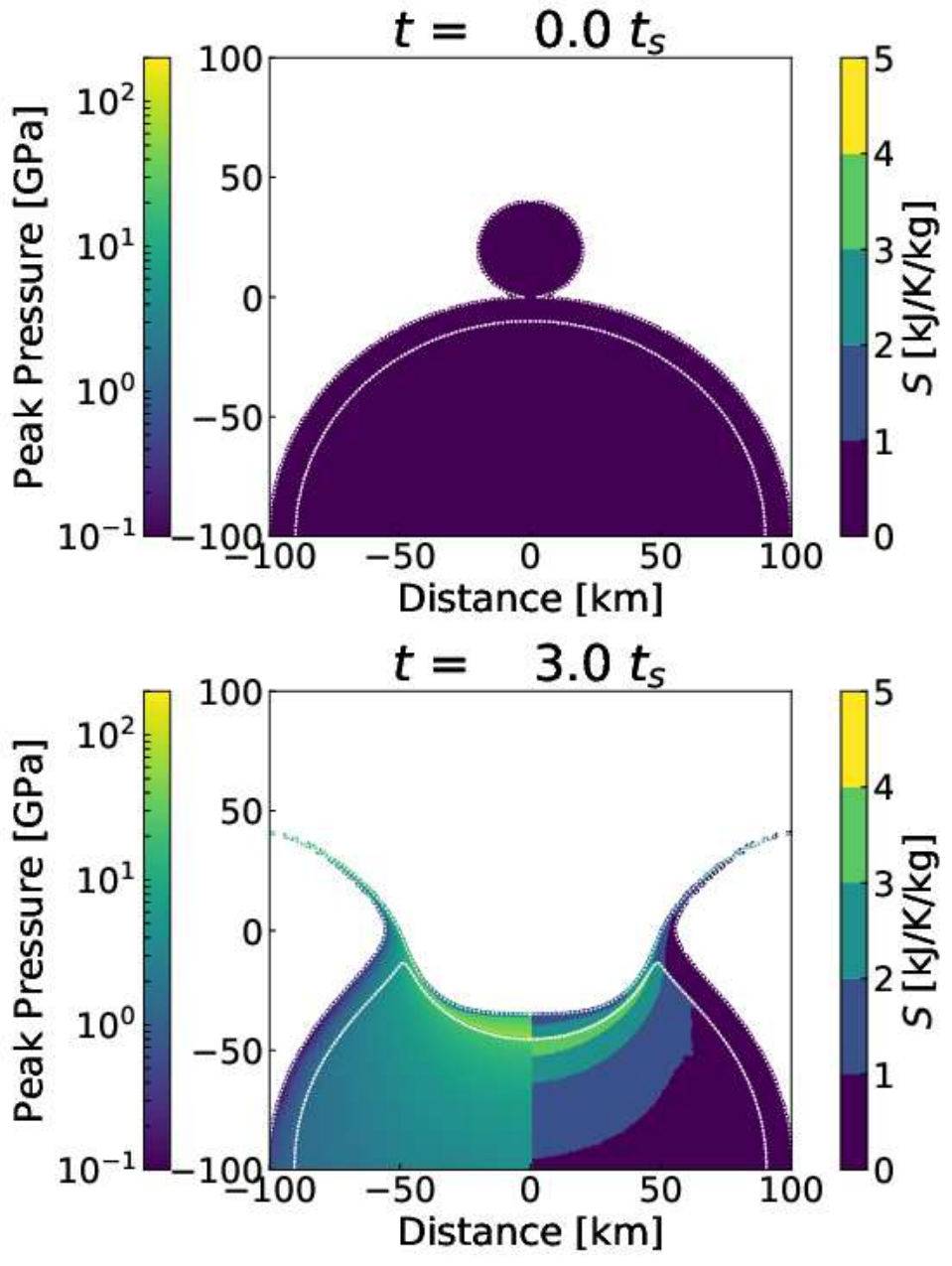}
\caption{Snapshot of a planetesimal collision between anhydrous impactor and hydrous target for the fiducial case
($R_{\rm imp}$ = 20 km, $R_{\rm tar}$ = 100 km and $R_{\rm hyd}$ = 90 km with $v_{\rm imp}$ = 5km/s) at 0 (top) and 3 (bottom) $t_s$. 
Color represents peak pressure on the left sides and the specific entropy at each position.
White dotted lines denote material boundary.
\label{fig:snap}}
\end{figure}

\begin{figure*}
\includegraphics[clip,scale=0.4]{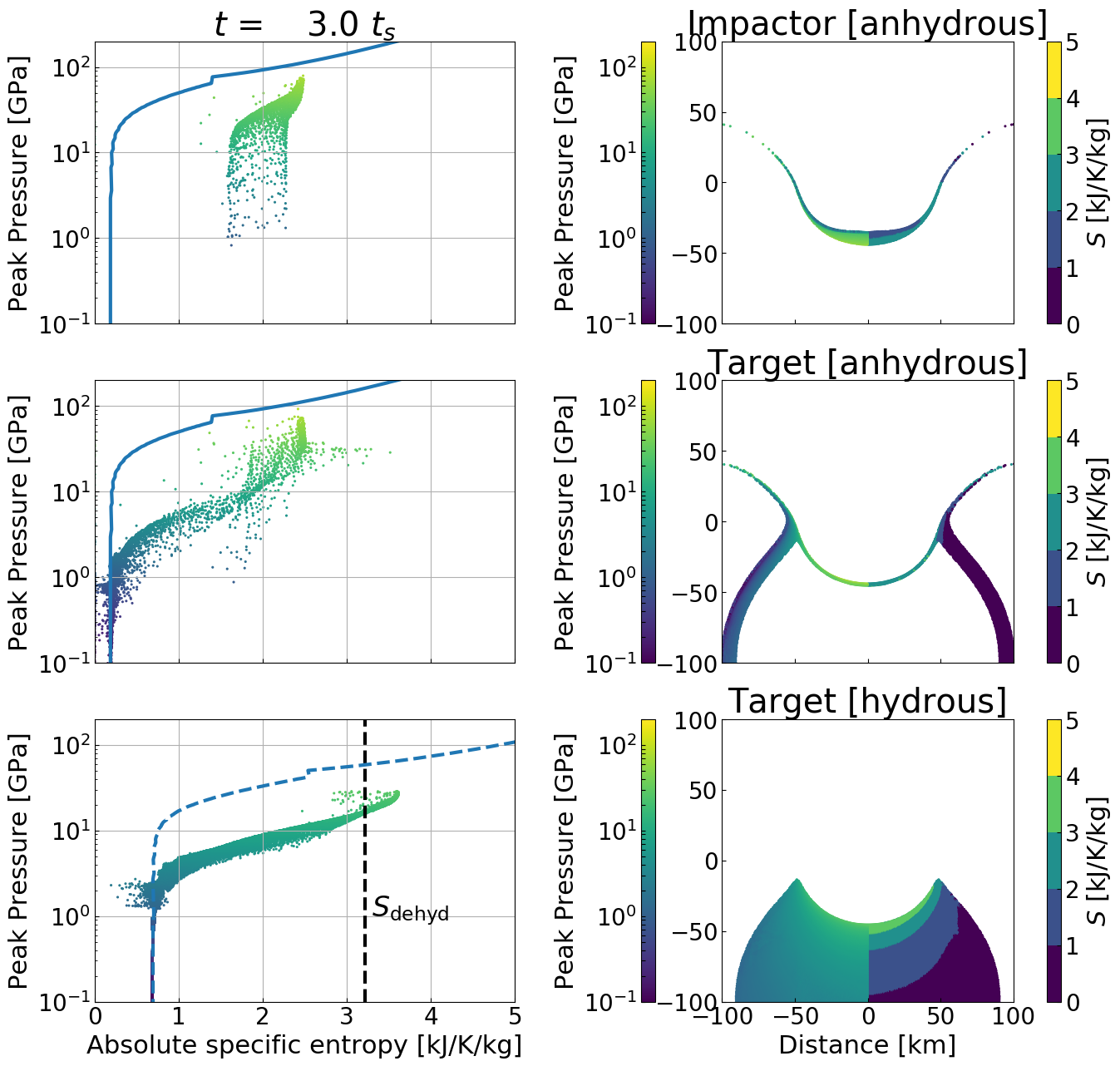}
\caption{Snapshot of a planetesimal collision between anhydrous impactor and hydrous target for the fiducial case
($R_{\rm imp}$ = 20 km, $R_{\rm tar}$ = 100 km and $R_{\rm hyd}$ = 90 km with $v_{\rm imp}$ = 5km/s) at 3 $t_s$. 
Left panels show peak pressure and absolute specific entropy with color contour of the peak pressure.
The Hugoniot curves (solid line for dunite and dashed one for serpentine) 
and the critical specific entropy of dehydration $S_{\rm dehyd}$ (vertical dashed line) are also shown.
Right panels depict the peak pressure on the left sides and the specific entropy at each position on the right sides.
Top panels denote the anhydrous impactor;
middle ones, the anhydrous layer in the target; 
and bottom ones, the hydrous core, respectively.
Dotted lines denote initial surface position of impactor and target,
and material boundary in the target.
\label{fig:fiducial}}
\end{figure*}

\begin{figure*}
\includegraphics[clip,scale=0.4]{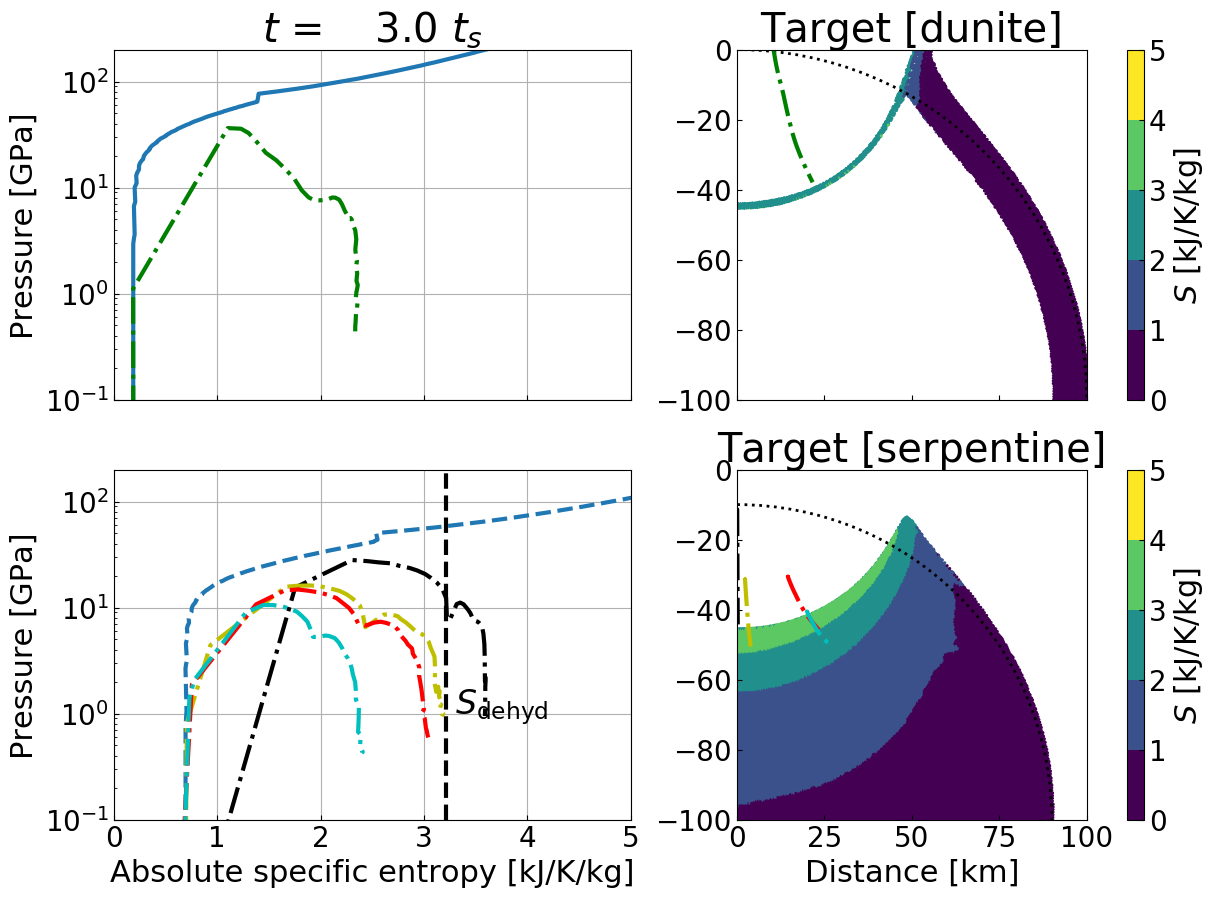}
\caption{Pressure and entropy change of selected tracer particles for the fiducial case.
Dash-dotted lines in each panel represent trajectories of tracer particles:
evolution of pressure and specific entropy on the left panels, 
and trajectory of position on right panels.
Left panels show pressure and absolute specific entropy. 
The Hugoniot curves (solid line for dunite and dashed one for serpentine) 
and the critical specific entropy of dehydration $S_{\rm dehyd}$ (vertical dashed line) are also shown.
Right panels depict the specific entropy at each position.
Top panels denote the anhydrous layer in the target,
and bottom ones the hydrous core of serpentine, respectively.
Dotted lines denote initial surface position of target,
and material boundary in the target.
\label{fig:spe}}
\end{figure*}
\begin{figure*}
\includegraphics[clip,scale=0.7]{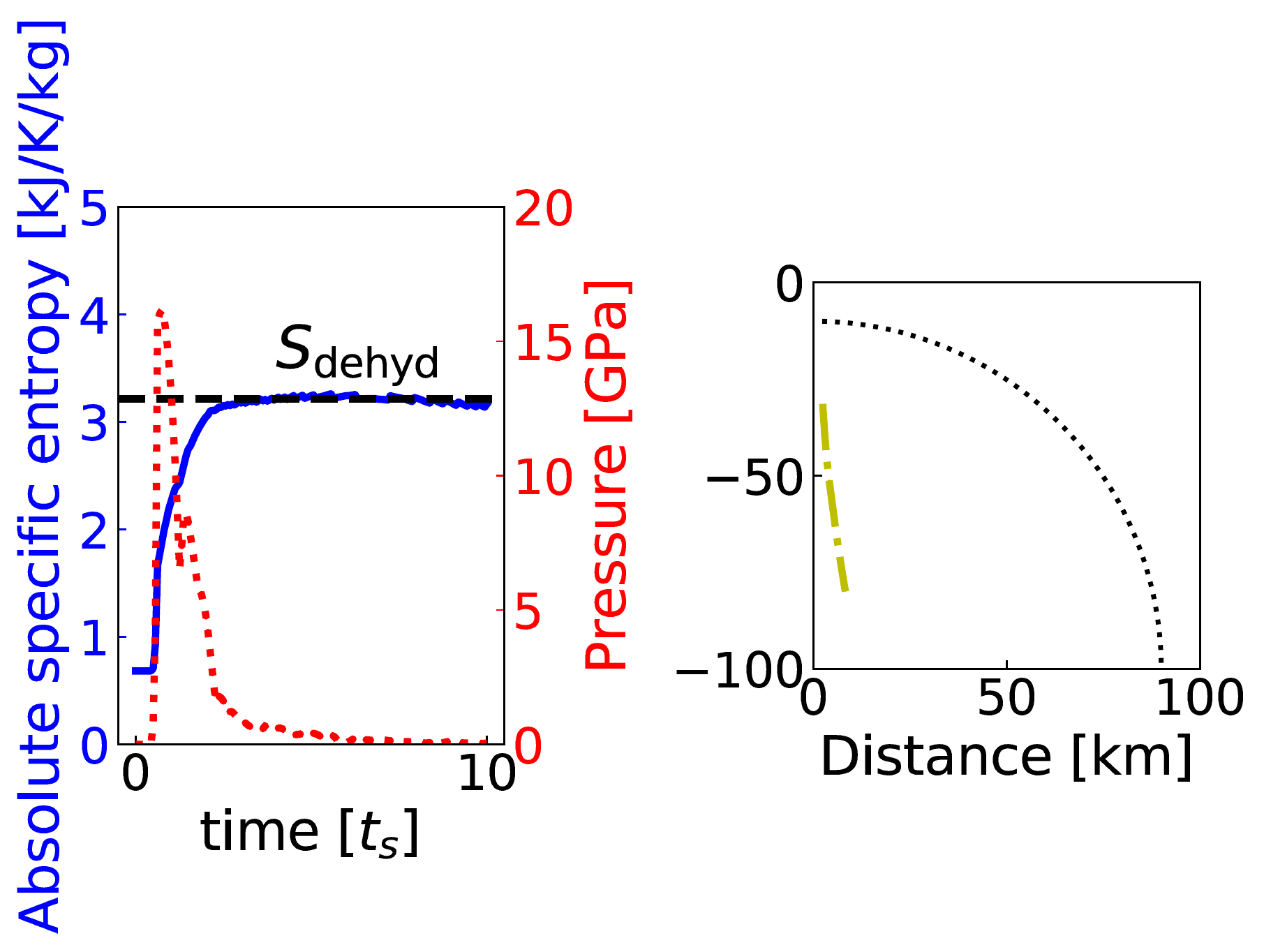}
\caption{Time evolution of the absolute specific entropy (blue solid line) and pressure (red dashed line) 
of a tracer particle in the hydrous core for the fiducial case.
The trajectory of this particle is shown in the right panel (same as yellow dash-dotted line in Figure \ref{fig:spe}).
\label{fig:trace}}
\end{figure*}

\begin{figure*}
\includegraphics[clip,scale=0.4]{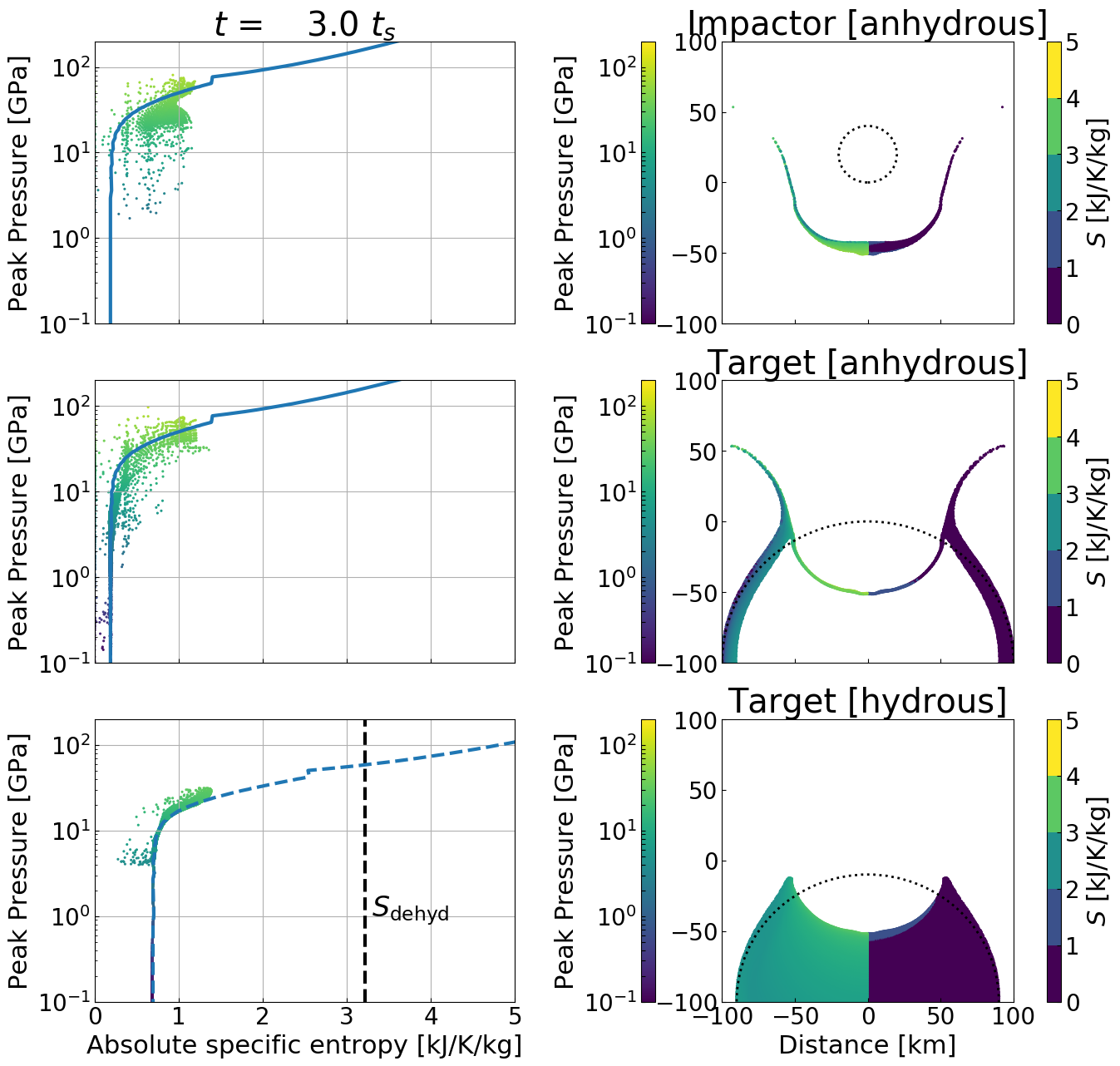}
\caption{Same as Figure \ref{fig:fiducial}, but the model does not include material strength.
\label{fig:sph}}
\end{figure*}
\begin{figure*}
\includegraphics[clip,scale=0.4]{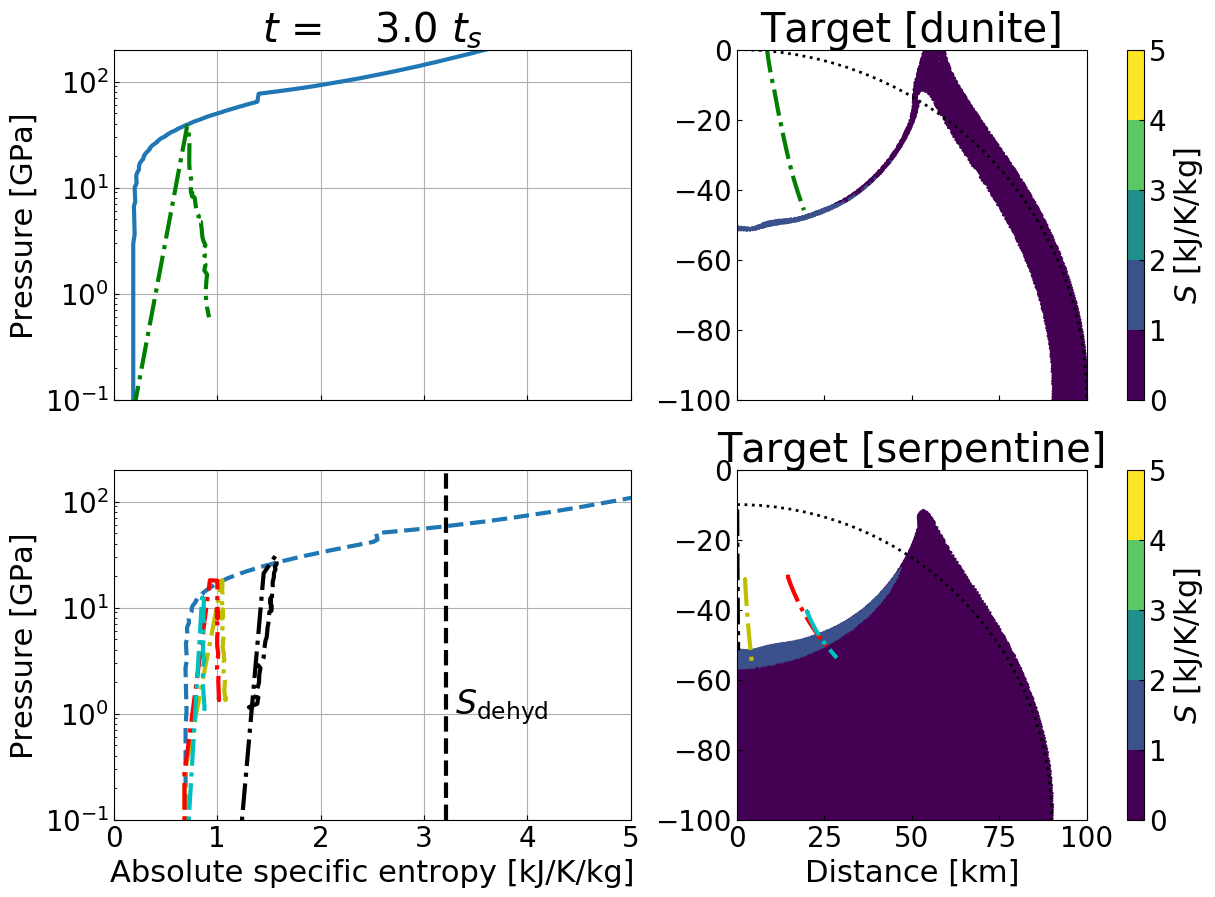}
\caption{Same as Figure \ref{fig:spe}, but the model does not include material strength.
\label{fig:speh}}
\end{figure*}
\begin{figure*}
\includegraphics[clip,scale=0.7]{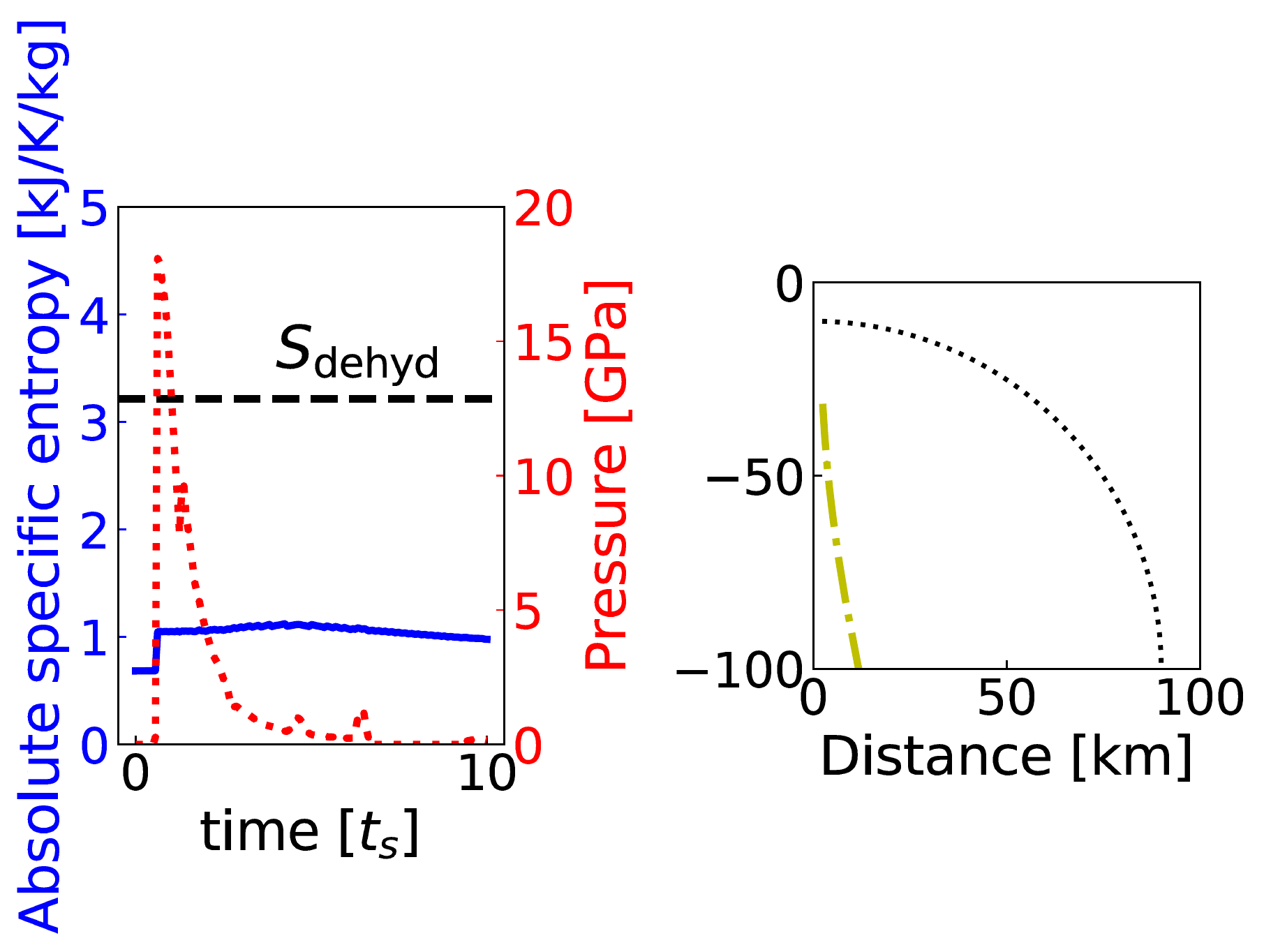}
\caption{Same as Figure \ref{fig:trace}, but the model does not include material strength.
\label{fig:traceh}}
\end{figure*}

Figure \ref{fig:fiducial} shows a snapshot for the fiducial case at t = 3.0 $t_s$, 
where $t_s$ is a characteristic time for projectile penetration ($t_s = 2R_{\rm imp}/v_{\rm imp}$),
and $t$ = 0 corresponds to the time when the two bodies make contact with each other.
Top panels denote results of a whole impactor; middle ones, the target's anhydrous layer; and bottom ones, the target's hydrous core, respectively.
The correlation of peak pressure ($P_{\rm peak}$) and absolute specific entropy ($S$) are shown on the left panels of Figure \ref{fig:fiducial}.
Lines in left panels denote Hugoniot curves, which are given by ANEOS in iSALE-2D; the solid one is from dunite and the dashed one is from serpentine.
As we take into account the strength model, the resultant peak pressure and entropy deviate from the Hugoniot curves as shown in \citet{Kurosawa:2018aa}.
This comes from the difference between the pressure and the longitudinal stress of elastoplastic media.
The Hugoniot relations go with the latter stress.
On the other hand, the peak pressure takes the former one, which is less than the latter stress by a factor of the Poisson ratio.
It is found that some hydrous materials in the hydrous core exceed $S_{\rm dehyd}$, which means that they experienced the dehydration reaction.
The amount of dehydrated materials is about 3 \% of the initial hydrous materials in this collision. 
Right panels in Figure \ref{fig:fiducial} depict the temporal pressure $P$ and the absolute specific entropy $S$ at each position.
We can see that dehydrated parts exist beneath the material boundary (bottom right panels of Figure \ref{fig:fiducial}).
Thus, hydrous materials can become dehydrated by planetesimal collisions.
The impactor is elongated by the impact and ejects materials from the collisional system (top right panels of Figure \ref{fig:fiducial}).
We also see the ejecta originating from the anhydrous layer (middle right panels of Figure \ref{fig:fiducial}).
Although the target planetesimal is excavated and reshaped by the impact,
their surface is still covered with anhydrous materials (see middle and bottom in right panels of Figure \ref{fig:fiducial}).

Lagrangian tracer particles are inserted in our numerical simulation.
The trajectories of some tracer particles are shown in Figure \ref{fig:spe}:
top panels denote results of the anhydrous layer and bottom ones are the hydrous core.
Temporal pressure ($P$) and absolute specific entropy ($S$) of selected tracer particles are shown on the left panels.
Their trajectories are shown on the right panels.
The entropy increases due to the passage of the shock wave, 
and continue to gradually increase during during pressure release.
The entropy of some parts in the hydrous core exceeds $S_{\rm dehyd}$.
In Figure \ref{fig:trace}, we plot the time evolution of $S$ and $P$ of a tracer particle which exceeds $S_{\rm dehyd}$.

It is worth seeing the effect of material strength by comparing to the model without material strength.
Figures \ref{fig:sph}, \ref{fig:speh}, and \ref{fig:traceh} show the result of a pure hydrodynamic case with the impact parameters same as in the fiducial case.
In Figure \ref{fig:sph}, most points follow the Hugoniot curves as expected.
Some points are above the Hugoniot curves, which come from close to the boundary, 
such as from between the impactor and target, and the target layer (dunite) and core (serpentine).
Some tracer particles deviate from the Hugoniot curves. 
We confirmed that this deviation occurs due to the following two reasons: 
unphysical change of the entropy during pressure release in iSALE code, 
and unphysical behavior at the material boundary and free surface.
Figure \ref{fig:sph} also shows that no tracer particles exceed $S_{\rm dehyd}$,
which is expected from simple analysis of the Hugoniot curve in Figure \ref{fig:vimp}.
We can confirm that the absolute specific entropy does not increase after the passage of the shock wave,
during adiabatic pressure release, in Figures \ref{fig:speh} and \ref{fig:traceh},
which means that dehydration does not occur under this impact condition.
These are also consistent with previous work \citep{Kurosawa:2018aa}.
Therefore, it is important to take into account the material strength for understanding the occurrence of the dehydration reaction.

\begin{figure}
\includegraphics[clip,scale=0.6]{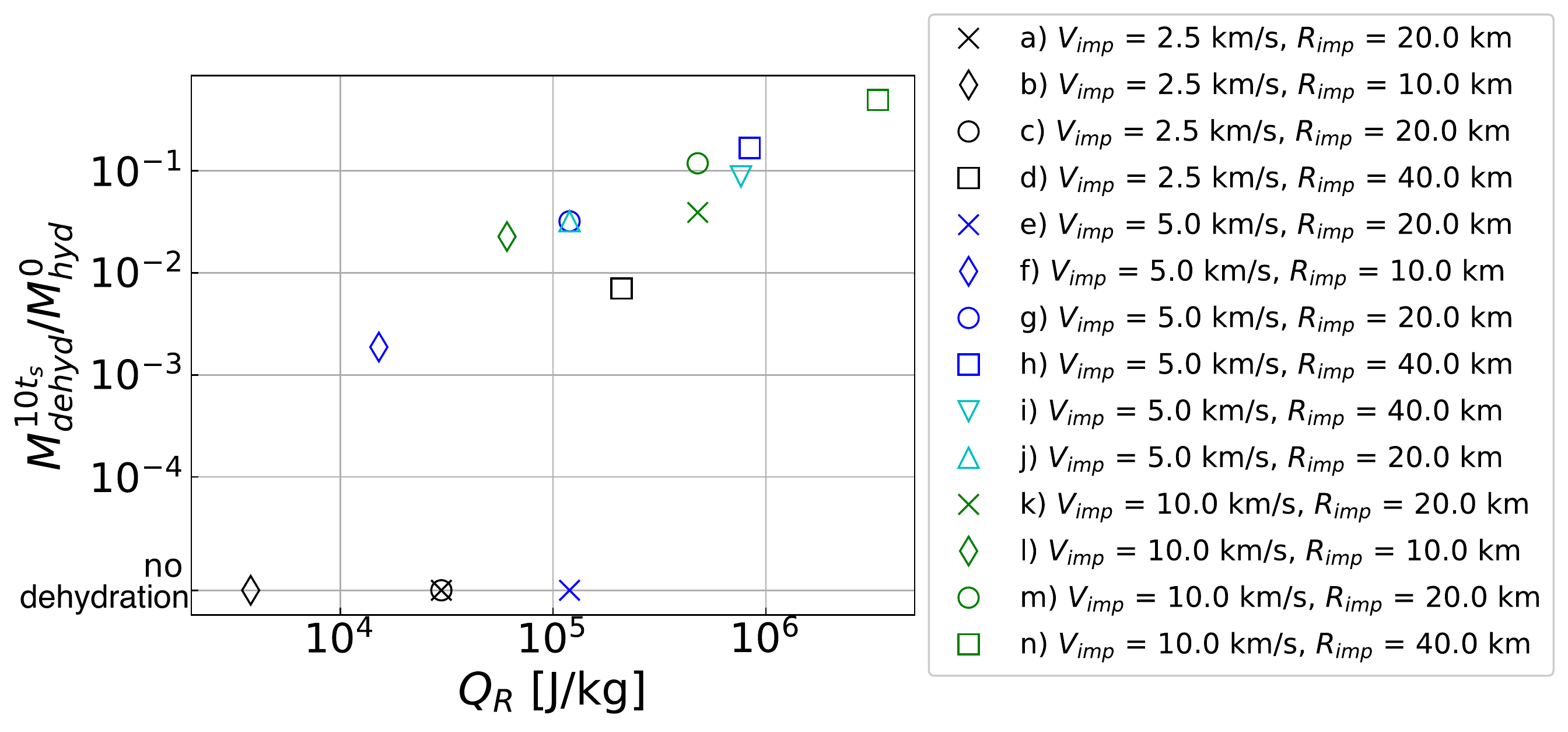}
\caption{The ratio of dehydrated mass to initial hydrated mass, $M^{10t_s}_{\rm dehyd}/M^0_{\rm hyd}$ as a function of specific impact energy $Q_R$.
Color represents each impact velocity and the symbol represents the conditions of each calculation.
Labels denote $V_{\rm imp}$ and $R_{\rm imp}$ of each symbol.
Cross symbols mean the models without material strength, and the other symbols are the models with material strength.
Upside-down triangle represents the model with the $R_{\rm hyd}$ = 70 km, and the other has $R_{\rm hyd}$ = 90 km.
Regular triangle represents the model with central gravity, and the other are the models without gravity.
Please note that in order to plot $M^{10t_s}_{\rm dehyd}/M^0_{\rm hyd}$ in one figure, log scale and $0$ values (labeled as "no dehydration") are shown at the same time.
\label{fig:md}}
\end{figure}

\subsection{Dependence on impactor size} \label{sec:rimp}
We investigate the effect of the impactor size on the amount of dehydration materials 
by changing the radius of the impactor from 20 km to 10 km and 40 km.
The other parameters are the same as those for the fiducial case.
Figure \ref{fig:md} shows the mass of dehydrated materials ($M^{10t_s}_{\rm dehyd}$) at 10 $t_s$ 
normalized by the initial mass of the hydrous core ($M^0_{\rm hyd} = 4/3\pi \rho_{\rm serp} R_{\rm hyd}^3$)
as a function of the specific impact energy ($Q_R$), 
where $Q_R$ is given as follows \citep{Leinhardt:2012aa,Genda:2017aa}:
\begin{equation}
Q_R = \left( \frac{1}{2}\frac{M_{\rm imp}M_{\rm tar}}{M_{\rm imp}+M_{\rm tar}} v_{\rm imp} \right) / (M_{\rm imp}+M_{\rm tar}),
\end{equation}
where $M_{\rm imp}$ and $M_{\rm tar}$ are the mass of the impactor and target, respectively.
The blue circle symbol in Figure \ref{fig:md} represents the result for the fiducial case ($R_{\rm imp}$ = 20 km), 
diamonds represent $R_{\rm imp}$ = 10 km, and squares represent $R_{\rm imp}$ = 40 km.
A larger impactor, which corresponds to larger $Q_R$, results in a larger amount of dehydration materials.
There is also a linear correlation in a log-log plot (which can be written as a power-law function in a linear plot) between $Q_R$ and $M^{10t_s}_{\rm dehyd}/M^0_{\rm hyd}$ at the fixed impact velocity of 5 km/s.

\subsection{Dependence on impact velocities} \label{sec:vimp}
Here, we can also see the dependence on impact velocities in Figure \ref{fig:md}.
Note we set other parameters to be the same as in the fiducial case: $R_{\rm imp}$ = 20 km and $R_{\rm hyd}$ = 90 km. 
The result of $M^{10t_s}_{\rm dehyd}/M^0_{\rm hyd}$ for $v_{\rm imp}$ = 10 km/s (green circle in Figure \ref{fig:md}) is larger than that for the fiducial case. 
This is because the $Q_R$ is larger for the fiducial case.
On the other hand, the result from $v_{\rm imp}$ = 2.5 km/s does not produce any dehydrated materials (see black circle in Figure \ref{fig:md}).
This means that $v_{\rm imp} \gtrsim$ 5 km/s with $R_{\rm imp}$ = 20 km can produce dehydrated materials in the hydrous core.

The dependence on impact velocity seen in Figure \ref{fig:vimp} can also be expected.
The critical velocity to produce the dehydrated materials is about 6 km/s.
Please note that this estimation can only be valid for the calculation of the pure hydrous case, i.e., without material strength \citep{Kurosawa:2018aa}.
Cross symbols in Figure \ref{fig:md} depict results without material strength.
In the case of $v_{\rm imp}$ = 5 km/s (blue cross in Figure \ref{fig:md}), the dehydration reaction does not occur (see also Figures \ref{fig:sph}, \ref{fig:speh}, and \ref{fig:traceh}).
The result of $v_{\rm imp}$ = 10 km/s (green cross in Figure \ref{fig:md}) could trigger the dehydration reaction.
However, the amount of dehydrated material $M^{10t_s}_{\rm dehyd}$ is less than what is seen with the case including material strength  (green circle in Figure \ref{fig:md}).
As a result, the material strength is an important factor in evaluating the onset and amount of dehydration in the hydrous core.

We also plot the result of $R_{\rm imp}$ = 10 km and 40 km with $v_{\rm imp}$ = 2.5 km/s and 10 km/s in Figure \ref{fig:md}.
As we see in \S \ref{sec:rimp}, a similar correlation can be seen for the cases producing dehydration materials during their collisions.

\subsection{Dependence on the size of the hydrous core} \label{sec:rhyd}
Changing the size of the hydrous core, we see the effect of the thickness of the initial anhydrous layer.
Figure \ref{fig:md} also shows the case of $R_{\rm hyd}$ = 70 km (blue triangle symbol) with $R_{\rm imp}$ = 40 km and $v_{\rm imp}$ = 5 km/s. 
The dehydration reaction also occurs in this case, 
but the amount of dehydrated materials ($M^{10t_s}_{\rm dehyd}/M^0_{\rm hyd}$ = 0.088) is less than it is in the case of $R_{\rm hyd}$ = 90 km (0.17).
This is because the latter case has a much thinner anhydrous layer than the case of $R_{\rm hyd}$ = 70 km.
Therefore, the size of the hydrous core (or the thickness of the initial anhydrous layer) is 
another important parameter to produce dehydration materials via planetesimal collisions.

\begin{figure}
\includegraphics[clip,scale=0.6]{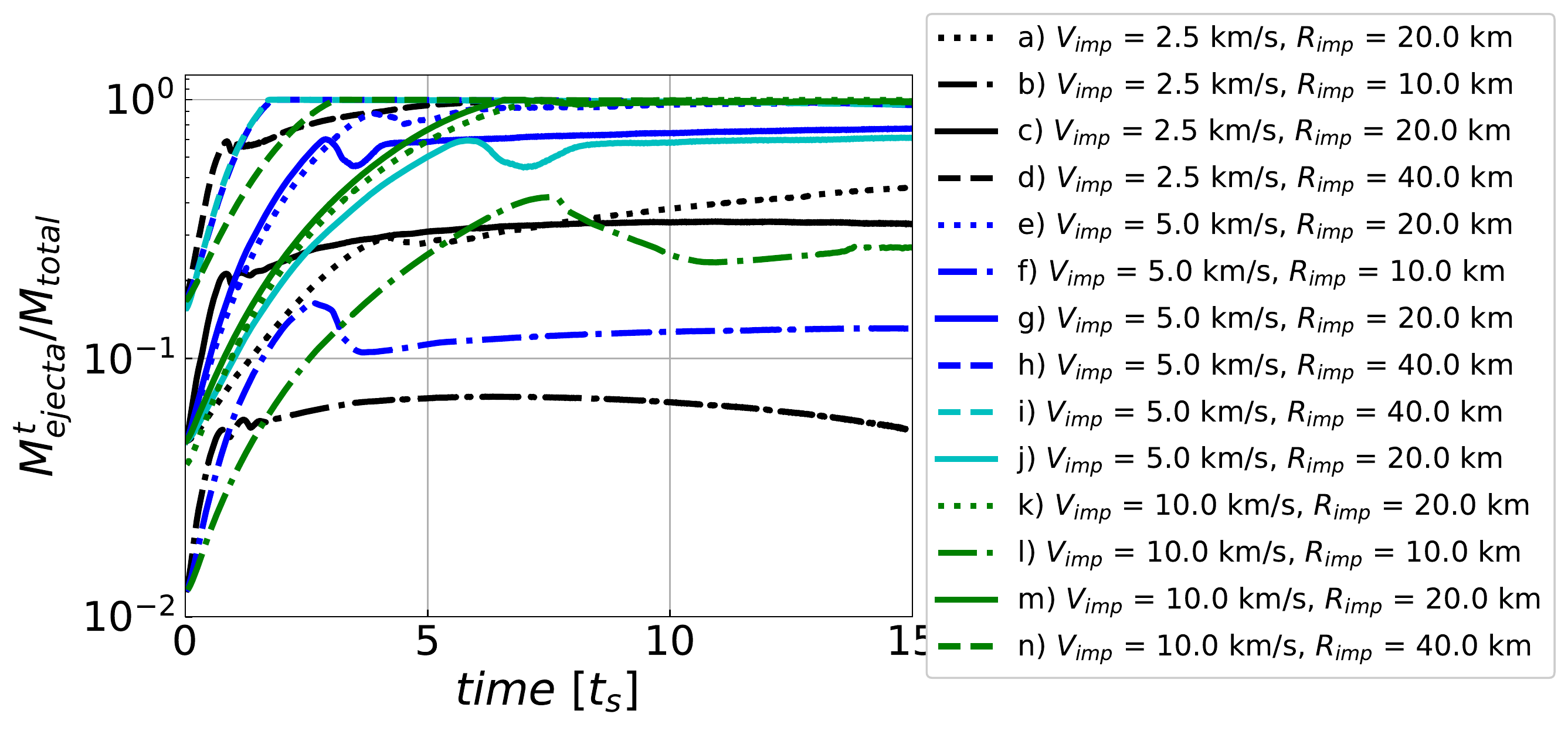}
\caption{The ratio of ejecta mass to initial mass, $M^{t}_{\rm ejecta}/M_{\rm total}$ as a function of time $t_s$.
\label{fig:mej}}
\end{figure}

\begin{figure}
\includegraphics[clip,scale=0.6]{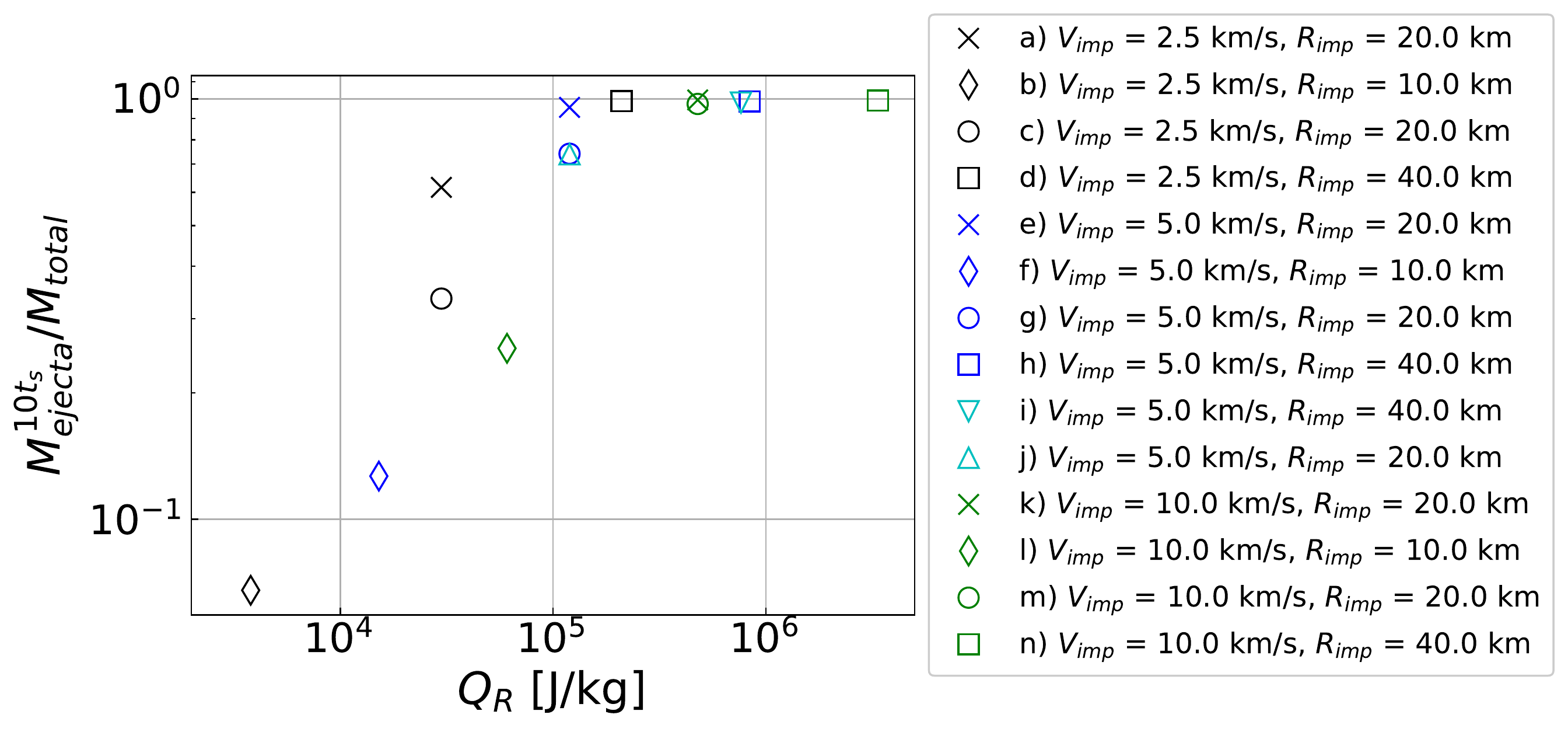}
\caption{The ratio of ejecta mass to initial mass, $M^{10t_s}_{\rm ejecta}/M_{\rm total}$ as a function of specific impact energy $Q_R$.
\label{fig:me}}
\end{figure}

\subsection{Fate of materials} \label{sec:fate}
In the previous sections, we present the amount of dehydration materials as the whole target.
However, high velocity collisions between planetesimals produce a lot of ejecta, and even cause disruption of the target.
Here we investigate the amount of materials for escaping ejecta and the remaining body.

\subsubsection{Ejecta from the system} \label{sec:ejecta}
The ejecta from the system (impactor and target planetesimals) and the remaining body are composed of hydrous and anhydrous materials.
Anhydrous materials originally come from anhydrous materials of the impactor and the target's surface layer, 
and dehydrated materials come from the target's core.
The ejecta escaping from the remaining body would accrete on to other planetesimals.
Therefore, hydrous ejecta might be the origin of hydrous materials on the surfaces of asteroids.

Here we define ejecta, such that the velocity of tracer particles $v$ exceeds the escape velocity of the system $v_{\rm esc}$,
\begin{equation}
v_{\rm esc} = \sqrt \frac{2G(M_{\rm imp} + M_{\rm tar})}{R_{\rm imp}+R_{\rm tar}}.
\end{equation}
Figure \ref{fig:mej} shows time evolutions of the mass of the ejecta ($M^{t}_{\rm ejecta}$) normalized by the total mass ($M_{\rm total} = M_{\rm imp} + M_{\rm tar}$).
Most of the lines are almost flat around 10 $t_s$.
Thus, we discuss the ejecta at 10 $t_s$ in the following.
Some lines (e.g., green dash-dotted) fluctuate. 
In this analysis, absolute values of the velocity for the tracer particles are considered, but their directions are not considered.
Therefore, some tracer particles with high velocity in the inner part of the target are classified into the ejecta at $< $10 $t_s$,
which would cause the fluctuation in Figure \ref{fig:mej}.
After 10 $t_s$, the estimated ejecta mass also relaxes.
Therefore, the relaxed ejecta mass would be robust.
It seems to be constant after 10 $t_s$, and it is reasonable to consider the ejecta at 10 $t_s$ even for this case.

We summarize $M^{10t_s}_{\rm ejecta}/M_{\rm total}$ in Figure \ref{fig:me} for all simulations as a function of $Q_R$.
As reported in many studies \citep[e.g.,][]{Benz:1999aa,Nakamura:2009aa},
the ejecta mass generally increases with $Q_R$.
The critical value of $Q_R$ for disruptive collisions has often been discussed \citep[e.g.,][]{Leinhardt:2012aa,Genda:2017aa}.
This critical value ($Q_{RD}^*$) is defined as the specific impact energy that half of the total mass lost after the collision.
When we extrapolate our results without material strength (see cross symbols in Figure \ref{fig:me}), 
the critical value is $2.0 \times 10^4$ [J/kg].
This is in good agreement with previous work \citep{Suetsugu:2018aa}, the value of which at $R_{\rm tar}$ = 100 km is about $1.8 \times 10^4$ [J/kg].
When we estimate $Q_{RD}^*$ from our results with material strength, it is about $10^5$ [J/kg].
This is about 5 times larger than results without material strength.
It indicates that a larger amount of energy is required to destroy half of the total mass when the material strength is considered.
This is also consistent with previous work \citep{Jutzi:2015aa}, which argues about the importance of friction of the target 
and finds that $Q_{RD}^*$ without material strength is 5 - 10 times larger than that with material strength.

\begin{figure*}
\includegraphics[clip,scale=0.4]{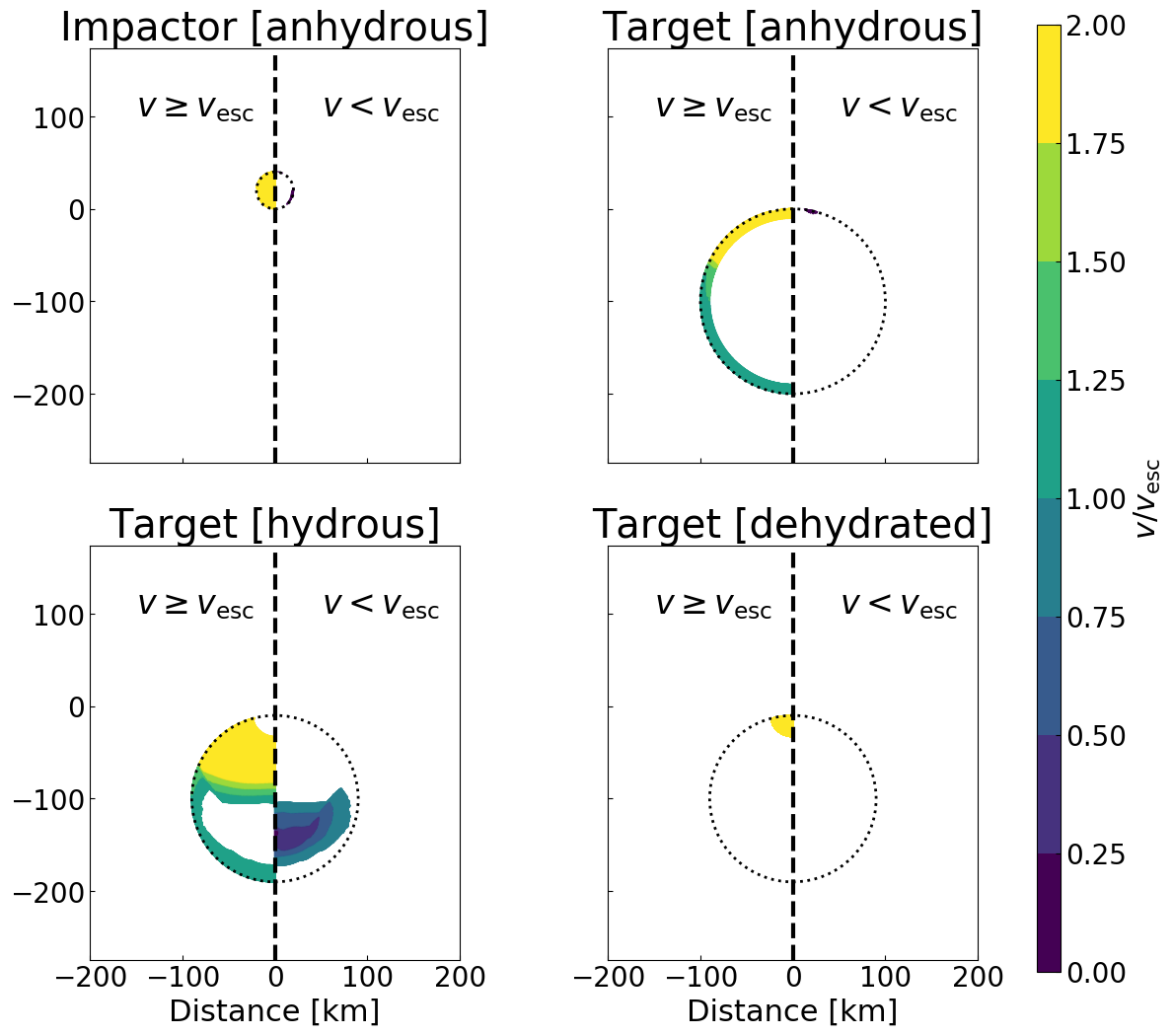}
\caption{Snapshot of planetesimal collision between anhydrous impactor and hydrous target for the fiducial case
($R_{\rm imp}$ = 20 km, $R_{\rm tar}$ = 100 km and $R_{\rm hyd}$ = 90 km with $v_{\rm imp}$ = 5km/s) at 10 $t_s$. 
Each panel represents velocity at each position as color contour:
top-left for impactor, top-right for anhydrous layer of target, 
bottom-left for hydrous core of target, and bottom-right for dehydrated part in target.
Tracer particles on right sides of each panel (positive side of horizontal axis) depict the position where the velocities are less than escape velocity of the system,
and ones on left sides (negative side of horizontal axis) have a velocity higher than the escape velocity.
Dotted lines denote initial surface position of impactor and target.
\label{fig:vesc}}
\end{figure*}

\subsubsection{Ratio of hydrous materials in ejecta and remnants}

Figure \ref{fig:vesc} shows the velocity of tracer particles normalized by the escape velocity, $v/v_{\rm esc}$, for the fiducial case.
The velocity of tracer particles at 10 $t_s$ is shown in their initial positions.
Those particles on the right sides of each panel (positive side of horizontal axis) depict the position where $v < v_{\rm esc}$:
tracer particles remain in the system.
On the other hand, the ones of $v \geq v_{\rm esc}$ are plotted on the left sides (negative side of horizontal axis):
ejecta from the system.
Most of the anhydrous materials in the impactor and target, and all dehydrated materials escape from the system.
A large amount of hydrous materials also escapes from the system without dehydration.
Since some of these ejected hydrous materials eventually accrete onto the other asteroids, 
it is possible that the hydrous materials on the surfaces of asteroids originated from inside other asteroids.

\begin{figure}
\includegraphics[clip,scale=0.6]{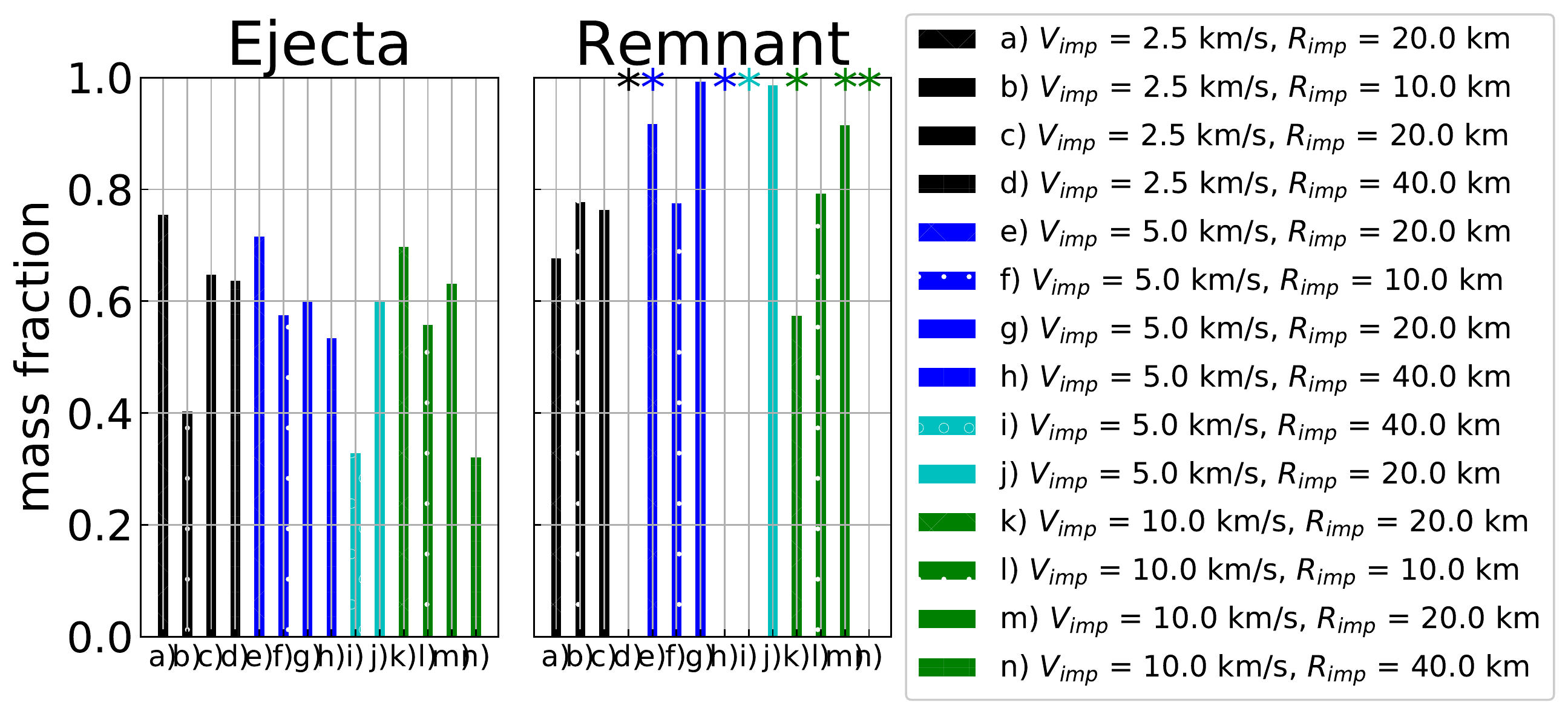}
\caption{The mass fraction of hydrous materials in ejecta (left) and remnant (right). 
\label{fig:mm}}
\end{figure}

The mass fraction of hydrous materials in ejecta and remnants are plotted in Figure \ref{fig:mm}.
In most cases, nearly half of the ejecta seems to be hydrous material.
On the contrary, the ratio of hydrous materials form a major component in the remnant. 
Note that the bar with the asterisk $*$ means that the mass of the remnant is too small or is at zero (see Figure \ref{fig:me}).
We can conclude that hydrous materials can avoid the dehydration reaction
and also be ejected from the system of planetesimal collisions.

\section{Discussions} \label{sec:dis}

\begin{figure}
\includegraphics[width=140mm]{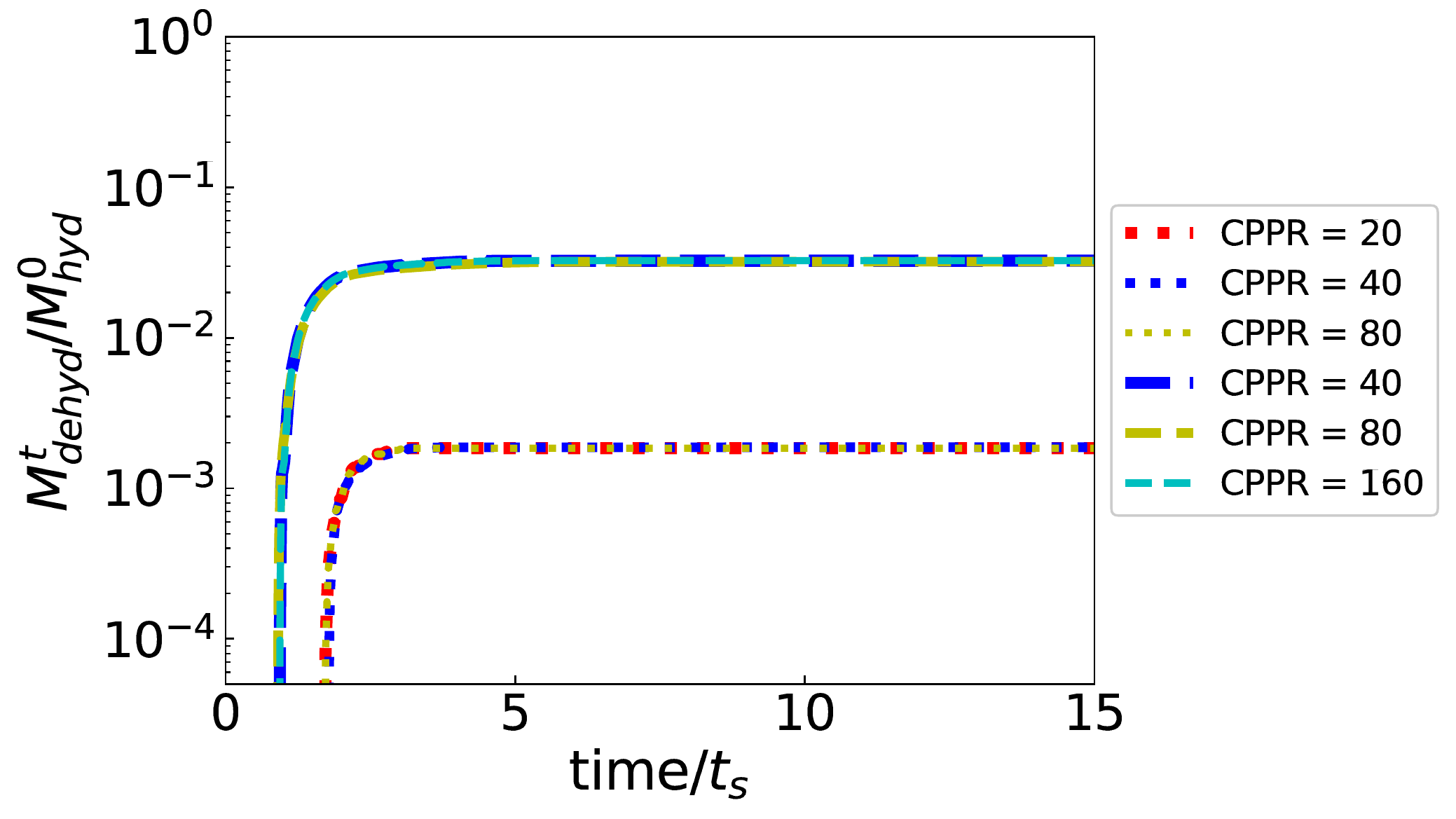}
\caption{Time evolution of the ratio of dehydrated mass to initial hydrated mass, $M^{10t_s}_{\rm dehyd}/M^0_{\rm hyd}$.
Each line corresponds to various cell per projectile radius (CPPR)  and various cases:
dashed lines denote fiducial cases ($R_{\rm imp}$ = 20 km, $R_{\rm tar}$ = 100 km, $R_{\rm hyd}$ = 90 km, and $v_{\rm imp}$ = 5km/s),
and dotted lines denote the case with $R_{\rm imp}$ = 10 km and the other parameters are the same as in the fiducial case.
\label{fig:cppr}}
\end{figure}

In the first half of this section, we discuss some numerical issues in our impact simulations.
In the previous section, we showed the results for the planetesimal collisions, whose numerical resolution is 250m per cell, 
which corresponds to 80 cells per projectile radius (CPPR) of $R_{\rm imp}$ = 20 km.
Figure \ref{fig:cppr} plots the time evolution of the ratio of dehydrated mass to initial hydrated mass, $M_{\rm dehyd}^t/M_{\rm hyd}^0$
for three different numerical resolutions.
The temporal values of $M^t_{\rm dehyd}/M_{\rm hyd}^0$ are almost the same between the case with the same $R_{\rm imp}$ but different CPPR.
Therefore, the simulations with CPPR = 40 for $R_{\rm imp}$ = 10 km and 80 for $R_{\rm imp}$ = 20 km are reasonable.

The mutual gravity of the system is not included in our simulations,
but this does not significantly change the ejecta mass in our simulation settings.
We define the materials whose ejection velocity ($v$) exceeds the escape velocity ($v_{\rm esc}$) as the ejecta.
The escape velocity of $M_{\rm tar}$ for the fiducial case is about 123 m/s.
We can estimate the effect of the gravity pull as $g \delta t$, 
where g is the gravitational acceleration ($g$ = 0.076 m/s$^2$) and $\delta t$ is the considered time duration.
This gravity pull reduces the velocity of tracer particles $v$.
When we take $\delta t$ as $t$ = 10 $t_s$ (= 80 s) for the fiducial case, $g \delta t$ is about 6 m/s.
Therefore, the effect of gravity pull is only 5 \% in the velocity and the amount of escaped materials would not change significantly.
We perform another simulation considering central gravity with the parameters of a fiducial case (model j in Table \ref{tab:par}).
Since the time evolution of the ejecta mass of a model with and without gravity differs (see g and j in Figure \ref{fig:mej}),
their total amount of ejecta mass differs $<$ 1 \% (see g and j in Figure \ref{fig:me}), which is on the order of our estimation.
The difference of mass fraction of hydrous materials in ejecta and remnant is also almost the same (see g and j in Figure \ref{fig:mm}).
Thus, we can say that our calculation without gravity is suitable.

We investigated only head-on collisions, but oblique collisions might change our results in some area.
As shown in \citet{Genda:2017aa}, the impact angle ($\theta$) is key for the mass of ejecta.
A head-on collision ($\theta$ = 0) can produce most ejecta, and the oblique collisions ($\theta >$ 0) have less ejecta for planetesimal-sized collisions.
Thus, the ejecta from the system should decrease when we consider oblique impacts.
However, an oblique impact may enhance the effect of frictional heating
because the leading side of the impact point experiences strong shear.
This might trigger the dehydration reaction.
In this sense, the dehydrated mass could increase for an oblique impact,
which should be tested in the future.

We also excluded the effect of porosity.
This is because we assume the planetesimal has a hydrous core that has experienced aqueous alteration.
Primitive planetesimals might contain pores filled with water (ice).
However, once the temperature increases in the planetesimals, aqueous alteration happens \citep[e.g.,][]{Grimm:1989aa,Wakita:2011aa}.
Then, the pores might be filled with products of aqueous alteration.
This is because hydrous minerals usually have a larger volume than anhydrous ones due to containing water.
If this is the case, a hydrous core would not have pores.
As long as our main target is the hydrous core, we can ignore the porosity in the hydrous core. 
There is, however, the possibility that the outer anhydrous layer of a target planetesimal would have some pores.
As shown in \citet{Davison:2010ab}, porous planetesimals can reach higher temperatures than non-porous bodies.
The outer anhydrous layer with pores can reach a higher temperature. 
Accordingly, the amount of dehydrated materials may increase.
On the other hand, shock waves effectively decays in the porous target.
In that case, the porous outer layer would be a buffer to cause less effective dehydration of the inner hydrous core.
Therefore, we need to pay careful attention to the porosity in the outer layer, which should also be considered in future work.

Next we discuss the implications of our numerical results on the observations of asteroids and analysis of meteorites.
As shown in Figure \ref{fig:md}, most hydrous minerals can avoid the dehydration reaction during planetesimal collisions
except for when there is very high impact velocity ($\sim$ 10 km/s), which is rare in the current main asteroid belt.
Depending on the impact conditions, some of the hydrous minerals can escape from the surviving planetesimal (see Figure \ref{fig:mm}).
Then, the hydrous ejecta orbits around the Sun.
Ejected materials would have various sizes.
When planetesimal collision occurs in the early solar system and produces small-sized grains, 
these grains interact with the remaining nebula gas, and accrete to the nearby planetesimals via pebble accretion \citep[e.g.,][]{Ormel:2010aa,Lambrechts:2012aa}.
Even if the ejected materials are large enough to be decoupled with the nebula gas, 
there is a chance for them to accrete onto a (nearby) body after long-term orbital evolution.
When re-accretion of these ejected materials occurs in the current main asteroid belt, other asteroids can accumulate hydrous minerals on their surface.
Therefore, the hydrous minerals on the surfaces of asteroids can originate from inside other planetesimals.

The same thing could have happened to the dehydrated ejecta.
The iron to magnesium ratio (Fe/Mg) becomes lower as the aqueous alteration proceeds \citep[e.g.,][]{Zolensky:1993aa}. 
The dehydrated minerals would keep this low Fe/Mg ratio of hydrous minerals.
Therefore, the Fe/Mg ratio would be key to distinguishing anhydrous minerals from dehydrated ones.
However, from ground observation of asteroids, we can not distinguish anhydrous minerals and dehydrated ones. 
This is because it is hard to measure the iron to magnesium ratio by spectral observation.
Thus, we cannot identify any dehydrated minerals on the surfaces of asteroids for now.
On the other hand, there are reports that dehydrated minerals are found in meteorites \citep[e.g.,][]{Nakamura:2005aa,Abreu:2013aa}.
The duration time of dehydration heating is found to be from hours to one thousand days, based on experiments of CM carbonaceous chondrite \citep{Nakato:2008aa}.
This duration time is much shorter than that of internal heating by short-lived radionuclides ($\sim$ Myr).
The peak metamorphic temperature of most carbonaceous chondrites cannot become high enough that the dehydration reaction can occur \citep{Scott:2014aa}.
Thus, external heating, such as that from an impact, would be a possible heating source for dehydrated minerals in CM chondrite.
Dehydrated materials in our simulations surely form via impact in a short timescale.
Therefore, dehydrated ejecta from the collision could become incorporated into other asteroids and be delivered to the Earth as meteorites after later impact events.

Except for extensively disruptive collisions, a target planetesimal can keep most of its original body.
This remaining body is mainly composed of hydrous materials (see Figure \ref{fig:mm}).
It is possible that hydrous materials could appear on the surface of the remaining body via excavation during planetesimal collisions.
If this is the case, this can be another origin of hydrous minerals on the current asteroids.
But if not, the anhydrous materials would cover the body.
The important point from our results is that, even if the body is covered with anhydrous materials,
a large amount of hydrous materials would exist inside the body.
Thus, if we can not find any hydrous minerals on the surface of an asteroid, it might still contain hydrous minerals inside it.
When we cannot detect hydrous minerals on the surface of Ryugu, which is the target of JAXA's on-going Hayabusa 2 mission,
it is possible that hydrous minerals are carried within Ryugu.
In this paper, we focus on the target planetesimal. 
However, there is a possibility that a hydrous impactor may bring hydrous minerals to the surfaces of asteroids. 
We will take it into account in the future.

\section{Conclusions} \label{sec:con}

Ground observations and exploration missions found 
that there are hydrous minerals on the surfaces of bodies in the current main asteroid belt \citep[e.g.,][]{Takir:2015aa,Russell:2015aa}.
Meteorites can also contain these minerals, 
and previous works indicate that such minerals would be formed via aqueous alteration in the planetesimals \citep[e.g.,][]{Grimm:1989aa,Davis:2014aa}.
On the other hand, the features of hydrous minerals can be eliminated by the dehydration reaction.
Temperature increases during planetesimal collision might cause this chemical reaction.
To understand the behavior of hydrous minerals during collisions, 
we numerically examine planetesimal collisions.
As a first step, we investigate head-on collisions between a hydrous target and anhydrous impactor. 
We examine the dehydrated materials when the entropy of hydrous ones exceeds a critical value based on experiments \citep{Lange:1982aa,Nozaki:2006aa,Nakato:2008aa}.
Our results show that the dehydration reaction could occur during the collisions.
However, the amount of dehydrated minerals is very small for typical impact conditions.
We also confirm that the material strength is important to consider the entropy change of minerals as shown in \citet{Kurosawa:2018aa}.
Most hydrous minerals can keep their composition during the collisions.
Moreover, the velocity of hydrous minerals exceeds the escape velocity of the collisional system.
Those hydrous minerals can escape from their host planetesimals and have a chance to accrete on to the surfaces of other asteroids.
Other hydrous minerals which have a lower velocity than the escape velocity remain their host planetesimals and can come out on to the surfaces of the bodies.
We suggest that the collisions can provide hydrous minerals to the surfaces of the asteroids.
The effects of oblique impacts and a hydrous impactor will be investigated in the future.

\section*{Acknowledgments}
We gratefully acknowledge the developers of iSALE-2D, including Gareth Collins, Kai W{\"u}nnemann, Dirk Elbeshausen, Tom Davison, Boris Ivanov and Jay Melosh,
and the developer of the pySALEPlot tool, Tom Davison.
We thank an anonymous referee and Tom Davison for their helpful comments and suggestions.
Numerical computations were partially carried out on the PC cluster at the Center for Computational Astrophysics, National Astronomical Observatory of Japan.
We also acknowledge the fruitful discussions at the 13th workshop on planetary impacts in 2017 and the 9th workshop on catastrophic disruption in 2018.
This work has been supported in part by a JSPS Grant-in-Aid for Scientific Research (17H06457).

\end{document}